# Strong coupling between an electron in a quantum dot circuit and a photon in a cavity

L.E Bruhat[1], T. Cubaynes[1], J.J. Viennot[2], M. C. Dartiailh[1], M.M. Desjardins[1], A. Cottet[1] and T. Kontos[1]*

[1] *Laboratoire Pierre Aigrain, Ecole Normale Supérieure-PSL Research University, CNRS, Université Pierre et Marie Curie-Sorbonne Universités, Université Paris Diderot-Sorbonne Paris Cité, 24 rue Lhomond, 75231 Paris Cedex 05, France*

[2] *JILA and Department of Physics, University of Colorado, Boulder, Colorado, 80309, USA*

*To whom correspondence should be addressed: kontos@lpa.ens.fr

Circuit quantum electrodynamics allows one to probe, manipulate and couple superconducting quantum bits using cavity photons at an exquisite level. One of its cornerstones is the possibility to achieve the strong coupling which allows one to hybridize coherently light and matter[1]. Its transposition to quantum dot circuits could offer the opportunity to use new degrees of freedom[2,3,4,5,6] such as individual charge or spin. However, the strong coupling of quantum dot circuits to cavity photons remains to be observed. Here, we demonstrate a hybrid superconductor-quantum dot circuit which realizes the strong coupling of an individual electronic excitation to microwave photons. We observe a vacuum Rabi splitting $2g \sim 10$ MHz which exceeds by a factor of 3 the linewidth of the hybridized light-matter states. Our findings open the path to ultra-long distance entanglement of quantum dot based qubits[7,8,9]. They could be adapted to many other circuit designs, shedding new light on the roadmap for scalability of quantum dot setups.

The engineering of electronic states in devices combining materials with different electronic properties is at the heart of many recent methods put forward for quantum information processing. One particularly promising venue is the coupling of superconductors to nanoconductors[10,11]. First, these setups allow one to control superconductivity at the nanoscale, as epitomized by the recent demonstration of the Andreev doublet[12]. Superconductors can also be used to probe and manipulate non-local superconducting correlations both in the non-topological[13,14] and in the topological regime[11,15].

Another wide class of devices which are sought for in quantum information processing is that of quantum dot circuits. In the quest of scalable architectures, the methods of cavity quantum electrodynamics have been implemented to couple the charge as well as the spin degrees of freedom to cavity photons, on a chip[2,3,4,5,6,16,17,19]. However, the strong coupling between light and matter, which arises when the light matter coupling strength exceeds the dissipation rates of both light and matter taken separately, remains to be demonstrated. Obtaining such a regime, signalled experimentally by a vacuum Rabi splitting in the cavity transmission spectrum, is pursued in many double dot quantum bit designs, including spin qubits[4,5], resonant exchange[8,9], or hybrid spin-charge qubits[6].

Our qubit design relies on the shaping of the spectrum of a double quantum dot by a superconducting reservoir as shown in figure 1d. We use carbon nanotube based double quantum dot circuits embedded in a high finesse superconducting microwave cavity[4,16,18]. The quantum dots made out of them can be attached to normal[16,17], ferromagnetic[4] or superconducting[13,19] reservoirs. This allows us to tailor a synthetic energy level structure from superconductor induced cotunnelling between the left and the right dot (see Methods). The latter property combined with a low charging energy allows us in this work to achieve the strong coupling regime between a single electronic excitation and a single photon.

Figure 1a, b and c show optical as well as scanning electron microscope pictures of one of our devices. Throughout the paper, we describe results obtained with two different devices (sample A and sample B) which had exactly the same layout. All the measurements have been carried out at about 18 mK. The layout of the double dots is presented in figure 1c. A single wall carbon nanotube is connected to a central superconducting finger and two outer non-superconducting electrodes (see Methods). A finger galvanically coupled to the central conductor of our cavity (in red) is attached to two top gates in a fork geometry. This coupling scheme is markedly different from the double-dot/cavity coupling schemes used so far in that context[2,3,4,17]. Instead of favouring a microwave modulation of the difference of the energy between the left and the right dot, the fork geometry shown in figure 1c favours the modulation of the sum of the left and right dot energies by microwave photons.

The interaction between our hybrid double quantum dot and the cavity photons is conveniently probed via the phase of the microwave signal transmitted through the cavity. Figure 2a and 2b display the phase contrasts for sample A and sample B tuned in gate regions with the largest signals. In figure 2a, one observes a "socket" shaped 0 phase contour line with a phase spanning from -40° to +40°. These features are similar for sample B presented in figure 2b, but the sign change as well as the socket shape are less obvious due to the strong absorption simultaneous to this large phase contrasts (see for example figure 4d). Similar to what has already been observed in double quantum dot setups[2,3,4,17], the sign change of the phase contrast signals a resonant interaction between a doublet involving one or two electrons on the double dot and the cavity photons. Specifically, the cavity provides a "cut" of the dispersion relation of the double dot like spectrum[4]. The contour line for 0° corresponds to the resonant condition: $\omega_{cav} \approx \omega_{DQD}$, where $\omega_{cav}$ is the cavity resonance frequency and $\omega_{DQD}$ is the hybrid double quantum dot resonance frequency. However, in contrast with the standard double quantum dot response, the resonance contour line is not along the zero-detuning line, but is distorted in the perpendicular direction. Using a microscopic theory of our hybrid superconductor-double quantum dot (see Methods), we can write the transition frequency of our device as:

$$\omega_{DQD} = \sqrt{(\epsilon_\delta^2 + 4t_b^2)Z(\epsilon_\delta, \epsilon_\Sigma)^2 + t(\epsilon_\delta, \epsilon_\Sigma)^2}$$

where $\epsilon_\delta = \epsilon_L - \epsilon_R$ and $\epsilon_\Sigma = \epsilon_L + \epsilon_R$ are respectively the detuning between the left(L) and the right(R) dot and the average energy of the two dots. The parameter $t_b$ is the bare hopping of the double quantum dot. The functions $Z(\epsilon_\delta, \epsilon_\Sigma)$ and $t(\epsilon_\delta, \epsilon_\Sigma)$ describe respectively a renormalization of detuning $\epsilon_\delta$ and of the L/R tunnelling. The socket shape of the transition line can be recast from the dependence of the functions Z and t on $\epsilon_\delta$ and $\epsilon_\Sigma$. As shown in figure 2c, the transition map expected from the theory (see Methods) as a function of $\epsilon_\delta$ and $\epsilon_\Sigma$, displayed in light brown is cut by the blue plane at the cavity frequency. This results naturally in a socket shaped transition frequency contour line. Finally, it is important to notice that the dependence of $t(\epsilon_\delta, \epsilon_\Sigma)$ upon $\epsilon_\Sigma$ yields a new light-matter coupling term for our device, along the lines of the original Loss and DiVincenzo proposal[20] and recent cavity-double quantum dot coupling proposals[6,8,9]. Indeed, in the Bloch sphere representation of figure 2d, the north and south poles are more along the detuning axis $\epsilon_\delta$ and the light–matter coupling indicated by a red arrow is mainly along the tunnel coupling axis, in stark contrast with the usual case for double quantum dots[2,3,4,17], where it is along $\epsilon_\delta$. The large phase contrast (about 100°) implies large distortions in the cavity transmission spectrum. We now focus on sample B which has the largest contrast.

The main result of our work is presented in figure 3a, top panel. Upon tuning sample B inside the large phase contrast zone of figure 2b, we observe a splitting in the cavity spectrum for an average number of photons $\bar{n}$ of about 1. This observation persists down to the lowest input power which corresponds to $\bar{n} \ll 1$ (see Supplementary Information (SI)). This is the hallmark of a vacuum Rabi splitting which indicates the strong coupling between our hybrid double quantum dot and the microwave cavity photons. Using a modelling based on two independent transitions (one very coherent, transition s, one less coherent, transition w), anticipating on the existence of a K/K' valley degree of freedom commonly observed in nanotubes, depicted in figure 3b, we are able to fit the data using a fully quantum numerical code (QuTip). As expected for a few level system, we are able to saturate the transitions and to recover the bare transmission of the cavity by injecting a large number of photons inside the cavity. In the present case, this saturation occurs for $\bar{n} \approx 100$. As shown in figure 3c, there is a continuous evolution from the vacuum Rabi splitting to the bare resonance from $\bar{n} \approx 0.1$ to $\bar{n} \approx 300$.

Why can we reach the strong coupling regime with an electronic excitation which is primarily charge like? Three main ingredients control the achievement of the strong coupling regime. The first is the light-matter coupling strength, the second is the linewidth of the atomic-like transition and the third is the line-width of the cavity. The latter is typically about 600 kHz and is therefore not a limiting factor since one can easily reach 10 MHz of charge-cavity coupling in most setups[2,3,4]. The main limitation of all the charge qubit like setups in cavity is the linewidth of the double dot transition which has been reported to be at least in the few 100 MHz range[17,22] so far. One important decoherence source explaining such a large linewidth is the background charge noise. Its spectral density $\langle \delta n^2 \rangle$ sets the dephasing rate $\Gamma_\varphi$ of the double dot, via the charging energy $E_C \sim e^2/C_\Sigma$, where $C_\Sigma$ is the total capacitance of the device[17,22,23,24], which can be expressed as:

$$\Gamma_\varphi \approx \frac{\partial \omega_{+-}}{\partial \epsilon} E_C \sqrt{\langle \delta n^2 \rangle} + \frac{1}{2} \frac{\partial^2 \omega_{+-}}{\partial \epsilon^2} E_C{}^2 \langle \delta n^2 \rangle + \cdots \qquad (1)$$

Above $\omega_{+-}$ is the frequency of the resonant transition controlled by energy $\epsilon$ which is a function of $\epsilon_\delta = \epsilon_L - \epsilon_R$ and $\epsilon_\Sigma = \epsilon_L + \epsilon_R$. The usual method to reduce the linewidth is to tune the system close to a sweep spot where the first order term of the above equation vanishes. However, this has turned out not enough to enter into the strong coupling regime. While new methods have recently appeared to mitigate charge noise[25], it is a priori very efficient to go towards small charging energy, in analogy with the transmon qubit[23]. The charging energy of sample A and B can simply be read-off from the transport stability diagram which is shown in figure 4b for sample A. Due to the fork-shaped top gates, our charging energy is 2 meV, about 10 times smaller than what we find typically for similar devices with a conventional top gate setting[4,17]. Since $\Gamma_\varphi \approx 400\ MHz$ in those conventional settings, a reduction of 10 of $E_C$ is expected to reduce $\Gamma_\varphi$ by a factor of 100, i.e. $\Gamma_\varphi \approx 4\ MHz$ . Importantly, this reduction also implies a decrease of the lever arm between the orbital energies $\epsilon_L$, $\epsilon_R$ of the dots and the cavity potentials. The coupling of photons through the variable $\epsilon_\delta$ used in former experiments[2,3,4,17] thus becomes too small to be exploited. However, our hopping $t(\epsilon_\delta, \epsilon_\Sigma)$ is tunable with the parameter $\epsilon_\Sigma$, which is naturally more strongly coupled to the cavity potential than $\epsilon_\delta$. This compensates the decrease of $E_C$ and gives us a large charge-photon coupling strength of about 10 MHz which allows us to reach the strong coupling regime.

These ingredients used to reach the strong electron-photon coupling are very generic and could be used in many other setups[6,8,9]. In our case, the microscopic origin of the tunable hopping is however related to the use of a hybrid superconductor double quantum dot setup, thanks to superconductor induced cotunneling processes. As shown in figure 4a, the presence of a superconductor renormalizes the hopping strength between the left and the right dot. The green dashed lines which indicate the $\omega_{cav} \approx \omega_{DQD}$ condition in figure 4c and 4d, show its dependence on the bias applied to the superconductor. This further confirms the tunability of our engineered L/R barrier as a function of $\epsilon_{\Sigma}$. Importantly, using local gates, one could also engineer a direct electrostatic control over the hopping strength.

Our findings open the path for entanglement of individual electron states over macroscopic distances[7,26]. The vacuum Rabi splitting of about 10MHz and the dot decay rate of about 2 MHz imply that we can engineer a fast coherent exchange flip-flop rate J. For two identical devices placed at each anti-node of the electrical field (separated by about 1 cm on our chip), the exchange rate would amount to:

$$2J = g_1 g_2 \left\{ \frac{1}{\Delta_1} + \frac{1}{\Delta_2} \right\} \approx 2.5 \; MHz$$

for vacuum Rabi splittings of $2g_1 = 2g_2 = 10 \; MHz$ and detunings $\Delta_1 = \Delta_2 = 20 \; MHz$. Combined with the recent proposal for Hamiltonian engineering in circuit QED[27], this would lead to an $\sqrt{iSWAP}$ operation in about 100 ns which would compare favourably with the recently developed two qubit gates based on Si quantum dots[28] and would be compatible with scalability. Our findings could also be instrumental for a direct study of Cooper pair splitting or Majorana bound states through microwave cavities[29,30].

**METHODS**

**Fabrication of the devices and measurement techniques.**

The microwave cavity is a Nb coplanar waveguide cavity with resonance frequency of about 6.635 GHz and a quality factor of about 16000. A 150nm thick Nb film is first evaporated on an RF Si substrate at rate of 1nm/s and a pressure of $10^{-9}$ mbar. The cavity is made subsequently using photolithography combined with reactive ion etching (SF6 process). Carbon nanotubes are grown with Chemical Vapor Deposition technique (CVD) at about 900°C using a methane process on a separate quartz substrate and stamped inside the cavity. The nanotubes are then localized. The fork top gate oxide is made using 3 evaporation steps of Al (2nm) followed each by an oxidation of 10 min under an $O_2$ pressure of 1 mbar. The Alox is covered by a Al(40nm)/Pd(20nm) layer. The nanotube is contacted with a central Pd(4nm)/Al(80nm) finger and two Pd(70nm) outer electrodes. The DC measurements are carried out using standard lock-in detection techniques with a modulation frequency of 137 Hz and an amplitude of 10 µV. The base temperature of the experiment is 18 mK. The microwave measurements are carried out using room temperature microwave amplifiers and a cryogenic amplifier (noise temperature about 5K) with a total gain of about 90 dB. We measure both quadratures of the transmitted microwave signal using an I-Q mixer and low frequency modulation at 2.7 kHz.

**Low energy Hamiltonian of a hybrid double quantum dot-superconductor device**

We present in this section explicit expressions of the functions $Z(\epsilon_\delta, \epsilon_\Sigma)$ and $t(\epsilon_\delta, \epsilon_\Sigma)$ mentionned in the main text. The full hamilonian of our hybrid double quantum dot-superconductor device reads:

$$\hat{H} = \hbar\,\omega_{cav}\,\hat{a}^\dagger\,\hat{a} + \epsilon_L\,\hat{n}_L + \epsilon_R\,\hat{n}_R + \hat{H}_{interaction} + \hat{H}_{tunnel\,dot/lead} + \sum_{k\sigma} E_k \gamma_{k\sigma}^\dagger \gamma_{k\sigma}$$

$$+ \left(g_L\,\hat{n}_L + g_R\,\hat{n}_R\right)\left(\hat{a} + \hat{a}^\dagger\right) + \hat{H}_{photon\,bath}$$

where $\omega_{cav}$ is the pulsation of the cavity and $\hat{a}^\dagger(\hat{a})$ the creation (annihilation) operators for the photon field, $\hat{n}_{L(R)}$ is the electron number operator for the L(R) dot, $\gamma_{k\sigma}^\dagger(\gamma_{k\sigma})$ are the creation(annihilation) Bogoliubov operators for the superconductor and $g_{L(R)}$ is the L(R) electron-photon coupling strength. The low energy spectrum of the system can be obtained by a Schrieffer-Wolf transformation corresponding to "tracing out" the superconducting quasiparticles and projecting on the relevant states, for example $\{|0,0\rangle, |0,1\rangle, |1,0\rangle, |1,1\rangle\}$.

Starting from the bare bonding/antibonding states $|+\rangle$, $|-\rangle$, we get in the case where the left(L)/right(R) tunnel rates to the superconductor $\Gamma_{SL}$ (resp. $\Gamma_{SR}$) are equal:

$$Z(\epsilon_\delta, \epsilon_\Sigma) \approx 1 + \pi\,t_{eh}^0\,\frac{2t_b}{\epsilon_\delta^2 + 4t_b^2}\,\frac{\epsilon_\Sigma - eV_S}{\Delta} \qquad (2)$$

$$t(\epsilon_\delta, \epsilon_\Sigma) \approx -\pi\,t_{eh}^0\,\frac{\epsilon_\delta}{\sqrt{\epsilon_\delta^2 + 4t_b^2}}\,\frac{\epsilon_\Sigma - eV_S}{\Delta} \qquad (3)$$

These functions depend on $t^0_{eh}$, the bare Cooper pair splitting amplitude, $t_b$ and $V_S$, the bias applied on the superconductor. Upon further analysis, we can account for the exact socket shape using the full expressions corresponding to (2) and (3) (not shown). For sample A, for example, this yields: $2t_b$=6.3GHz, $t^0_{eh}$=400 MHz, $\Gamma_{SL}$=400 MHz and $\Gamma_{SR}$=800 MHz.

The electron-photon coupling strength is controlled by the sum $g_L + g_R$ which can easily be of the order of 100 MHz, as shown for example in reference [17] of the main text, which is a large magnitude. This explains why the "second order" term obtained from the superconductor can yield a sizable coupling strength.

1. Wallraff, A., Schuster, D.I., Blais, A., et al, Strong coupling of a single photon to a superconducting qubit using circuit quantum electrodynamics. *Nature* **431**, 162 (2004).

2. Frey, T., Leek, P. J., Beck, M., Blais, A., et al., Dipole Coupling of a Double Quantum Dot to a Microwave Resonator, *Phys. Rev. Lett.* **108**, 046807 (2012).

3. Petersson, K. D., McFaul, L. W., Schroer, M. D., Jung, M., et al., Circuit quantum electrodynamics with a spin qubit, *Nature* **490**, 380–383 (2012).

4. Viennot, J.J., Dartiailh, M.C., Cottet, A., and Kontos , T., Coherent coupling of a single spin to microwave cavity photons, *Science* **349**, 408 (2015).

5. Tosi, G., Mohiyaddin, F. A., Huebl, H., Morello, A., Circuit-quantum electrodynamics with direct magnetic coupling to single-atom spin qubits in isotopically enriched $^{28}$Si, *AIP Advances* **4**, 087122 (2014).

6. Friesen, M., Eriksson, M. A., Coppersmith, S. N., A decoherence-free subspace for charge: the quadrupole qubit, arXiv:160501797 (2016)

7. Burkard, G., Imamoglu, A., Ultra-long-distance interaction between spin qubits, *Phys. Rev. B* **74**, 041307 (2006).

8. Russ, M. and Burkard, G., Long distance coupling of resonant exchange qubits, *Phys. Rev. B* **92**, 205412 (2015)

9. Srinivasa, V., Taylor, J. M., Tahan, C., Entangling distant resonant exchange qubits via circuit quantum electrodynamics, arXiv:1603.04829 (2016)

10. Janvier, C., Tosi, L., Bretheau, L., Girit, Ç.Ö., et al., Coherent manipulation of Andreev states in superconducting atomic contacts, *Science* **349**, 1199 (2015).

11. Albrecht, S. M., Higginbotham, A. P., Madsen, M., et al., Exponential protection of zero modes in Majorana islands, *Nature* **531**, 206 (2016).

12. Bretheau, L., Girit, Ç. Ö., Pothier, H., Esteve, D., Urbina, C., Exciting Andreev pairs in a superconducting atomic contact, *Nature* **499**, 312 (2013).

13. Herrmann, L.G., Portier, F., Roche, P., et al., Carbon Nanotubes as Cooper-Pair Beam Splitters, *Phys. Rev. Lett.* **104**,115414 (2010).

14. Hofstetter, L., Csonka, S., Nygard, J., and Schönenberger, C., Cooper pair splitter realized in a two-quantum-dot Y-junction, *Nature* **461**, 960 (2009).

15. Mourik, V., Zuo, K., Frolov, S. M., et al., Signatures of Majorana Fermions in Hybrid Superconductor-Semiconductor Nanowire Devices, *Science* **336**, 1003 (2012).

16. Delbecq, M.R., Schmitt, V., Parmentier, F.D., et al., Coupling a Quantum Dot, Fermionic Leads, and a Microwave Cavity on a Chip, *Phys. Rev. Lett.* **107**, 256804 (2011).

17. Viennot, J. J., Delbecq, M. R., Dartiailh, M. C., Cottet, A., Kontos, T., Out-of-equilibrium charge dynamics in a hybrid circuit quantum electrodynamics architecture, *Phys. Rev. B* **89**, 165404 (2014).

18. Viennot, J. J., Palomo, J., Kontos, T., Stamping single wall nanotubes for circuit quantum electrodynamics, *Appl. Phys. Lett.* **104**, 113108 (2014).

19. Bruhat, L.E., Viennot, J.J., Dartiailh, M.C., et al., Cavity Photons as a Probe for Charge Relaxation Resistance and Photon Emission in a Quantum Dot Coupled to Normal and Superconducting Continua, *Phys. Rev. X* **6**, 021014 (2016).

20. Loss, D., DiVincenzo, D. P., Quantum computation with quantum dots, *Phys. Rev. A* **57**, 120 (1998).

21. Stockklauser, A., Maisi, V. F., Basset, J., et al., Microwave Emission from Hybridized States in a Semiconductor Charge Qubit, *Phys. Rev. Lett.* **115**, 046802 (2015).

22. Cottet, A., Vion, D., Joyez, P., et al., Implementation of a combined charge-phase quantum bit in a superconducting circuit, *Physica C* **367**, 197 (2002).

23. Koch, J., Yu, T. M., Gambetta, J., et al., Charge-insensitive qubit design derived from the Cooper pair box, *Phys. Rev. A* **76**, 042319 (2007).

24. Cottet, A., Kontos, T., Spin Quantum Bit with Ferromagnetic Contacts for Circuit QED, *Phys. Rev. Lett.* **105**, 160502 (2010).

25. Russ, M., Ginzel, F. and Burkard, G., Coupling of three-spin qubits to their electric environment, arXiv:1607.02351 (2016)

26. Majer, J., Chow, J. M., Gambetta, J. M., et al., Coupling superconducting qubits via a cavity bus, *Nature* **449**, 443 (2007).

27. Beaudoin, B., Blais, A., Coish, W. A., Hamiltonian engineering for robust quantum state transfer and qubit readout in cavity QED, arXiv:1602.05090 (2016)

28. Veldhorst, M., Yang, C.H., Hwang, J.C.C., et al., A two-qubit logic gate in silicon, *Nature* **526**, 410 (2015).

29. Cottet, A., Kontos, T. and Levy Yeyati, A., Subradiant Split Cooper Pairs, *Phys. Rev. Lett.* **108**, 166803 (2012).

30. Cottet, A., Kontos, T. and Douçot, B., Squeezing light with Majorana fermions, *Phys. Rev. B* **88**, 195415 (2013).

**Supplementary Information** Supplementary Information accompanies the paper on www.nature.com/nature.

**Acknowledgements** We are indebted with J.M. Raimond, C. Schönenberger and A. Baumgartner for discussions. We acknowledge technical support from J. Palomo, M. Rosticher and A. Denis. The devices have been made within the consortium Salle Blanche Paris Centre. This work is supported by ERC Starting Grant CIRQYS.

**Authors contributions**

**Competing financial interests** The authors declare no competing financial interests.

**Correspondence** Correspondence and requests for materials should be addressed to T.K. : kontos@lpa.ens.fr

## Figure 1 | Superconductor double quantum dot-cavity QED setup

**a.** Optical photograph of the layout of our cavity QED architecture on a large scale. **b.** and **c.** SEM micrographs of our devices on two different scales in false colours. The 'fork' coupling gate is coloured in red. The superconducting electrode is coloured in orange. The normal (non-superconducting) electrodes are coloured blue. The gates are coloured in green. **d.** Circuit diagram of our hybrid double quantum dot highlighting the symmetric coupling scheme between the two dots and the resonator in red.

## Figure 2 | Coupling to the microwave field induced by tunneling

**a.** and **b.** Microwave phase contrasts maps for sample A and sample B as a function of the gate electrodes. The sign changes demonstrate resonant interaction between the hybrid double quantum dot and the cavity photons. The elongated 0 phase line demonstrates the dependence of hopping with $\varepsilon_\Sigma$. **c.** Diagram of the transition map of the double quantum dot intersecting with the cavity resonance frequency. This results in the phase contrast maps of panels **a.** and **b. d.** Bloch sphere diagram depicting the active states of our hybrid double quantum dot and the tunable hoping strength. This symmetric coupling scheme is crucial for the strong electron-photon coupling.

## Figure 3 | Vacuum Rabi splitting.

**a.** Top panel: Vacuum Rabi splitting for sample B with n~1 photon. Bottom panel: Saturation of the mode splitting for a large number of photons. The open blue circles are the data points and the black solid line is the theory.  **b.** Level structures explaining the strong coupling and its power dependence. The K, K' labels indicate the valley degree of freedom arising from the band structure of carbon nanotubes.  **c.** Power dependence of the mode splitting showing the gradual saturation of the coherent transition. Each cut can be fitted using the fully quantum light-matter interaction theory (QuTip).

**Figure 4 | Superconductor induced cotunneling.**

**a.** Cotunneling scheme accounting for the renormalization of the hopping between the left and the right quantum dot. **b.** Colourscale map of the current flowing through the left (L) contact as a function of bias voltage ($V_{sd}$) and the detuning $V_{\delta}$ for sample B. From this map, we read-off a superconducting gap~150 $\mu$V. **c.** and **d.** Colourscale maps of the amplitude of the transmitted microwave field in the bias-gate plane. The green dashed lines highlight the $\omega_{cav} \approx \omega_{DQD}$ absorption line.

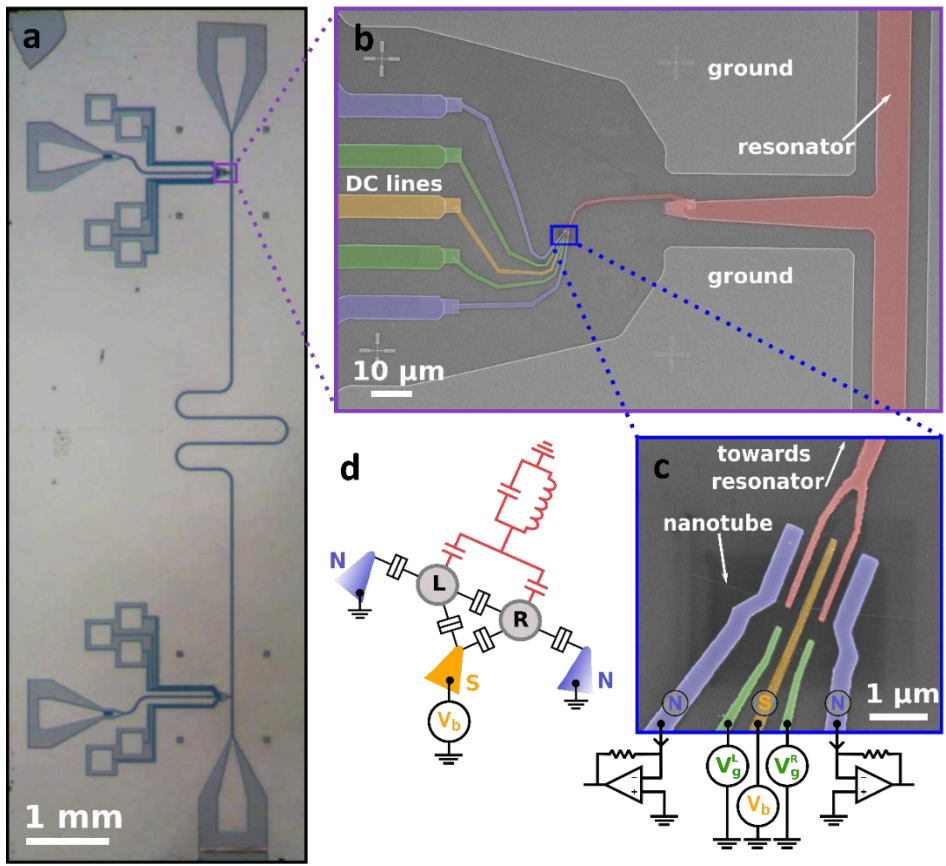

Fig1. Bruhat et al.

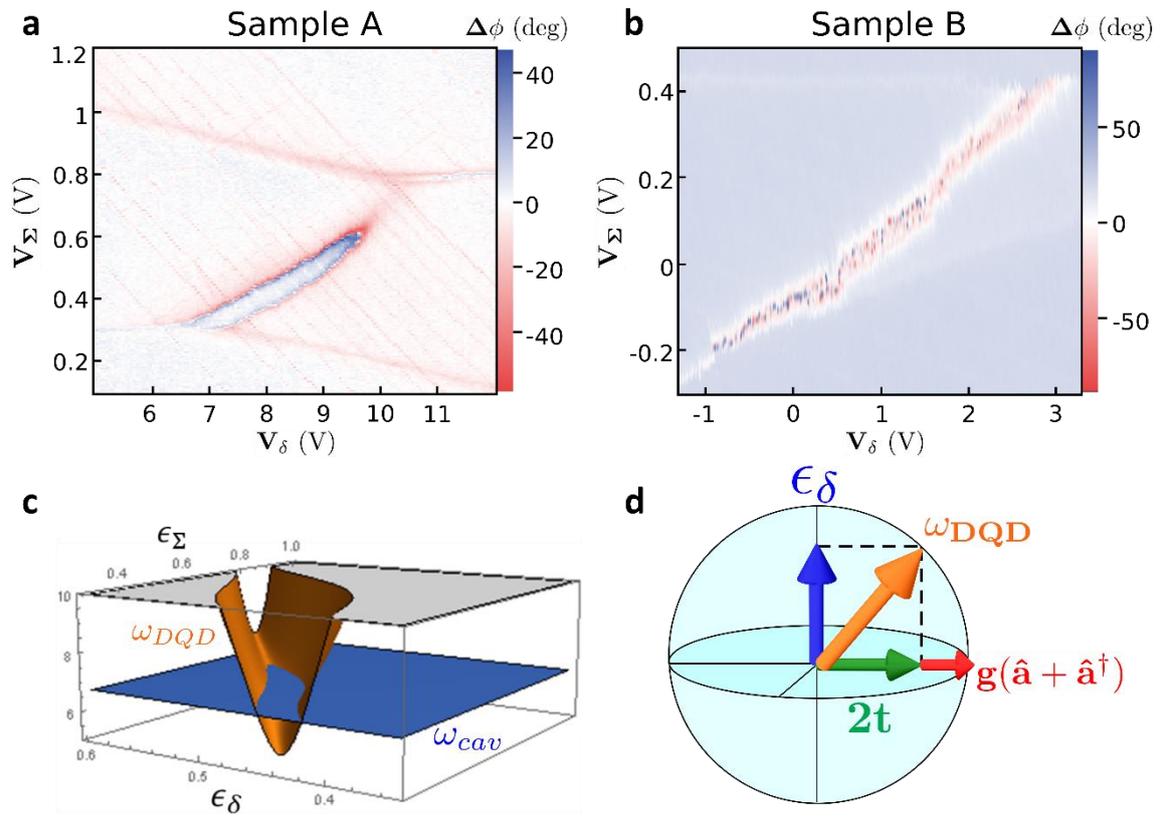

Fig. 2 Bruhat et al.

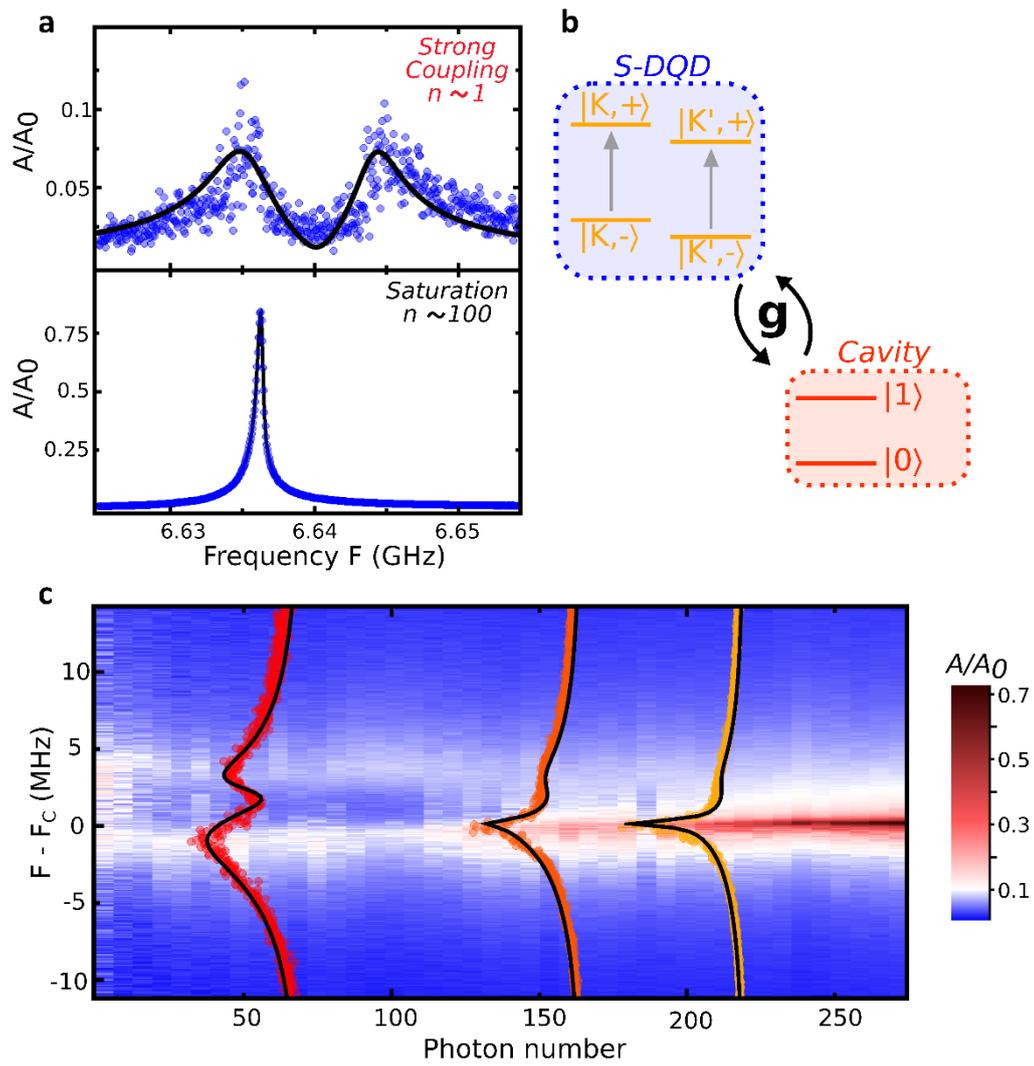

Fig. 3 Bruhat et al.

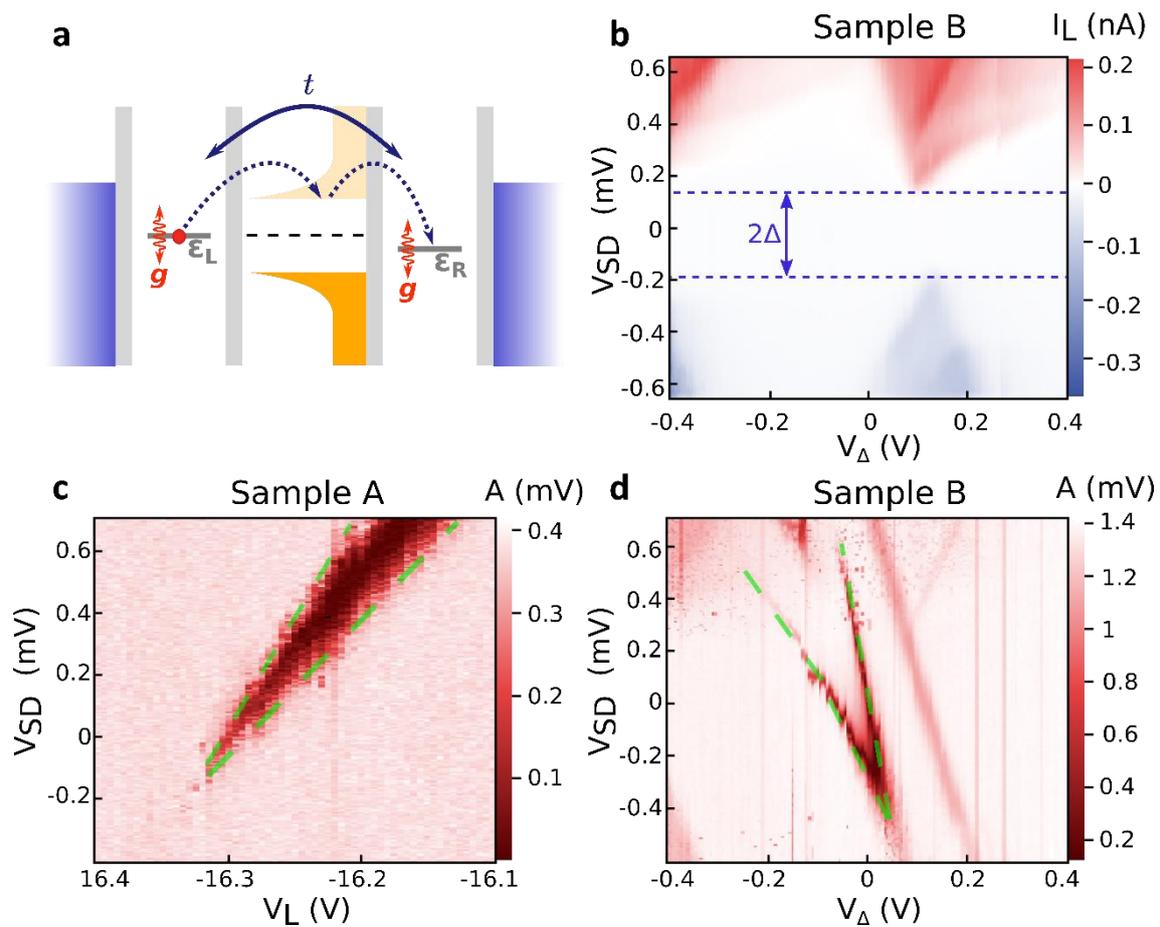

Fig. 4 Bruhat et al.